\documentclass[%
 reprint,
 amsmath,amssymb,
 aps,
 pra,
]{revtex4-2}

\usepackage{graphicx}
\usepackage{dcolumn}
\usepackage{bm}
\usepackage{amsthm}
\usepackage{amsmath,amssymb}
\usepackage{mathtools}
\usepackage{braket}
\usepackage{physics}
\usepackage{qcircuit}
\usepackage{tikz-cd}
\usepackage{quiver}
\usepackage{dsfont}
\usepackage{xcolor}
\usepackage{soul}
\usepackage{subcaption}
\usepackage{booktabs}
\usepackage{algorithm}
\usepackage{algpseudocode}
\usepackage{lipsum}
\usepackage[utf8]{inputenc}
\usepackage[colorlinks=true, allcolors=blue]{hyperref}

\usepackage[utf8]{inputenc}
\usepackage[justification=raggedright,singlelinecheck=false]{caption}

\usepackage{etoolbox}    

\usepackage{caption}
\usepackage{xcolor}
\newif\ifmarked
\markedfalse   

\ifmarked
  
  \newenvironment{review}{\color{red}}{}
\else
  
  \newenvironment{review}{}{}
\fi

\begin{document}

\preprint{APS/123-QED}

\title{Thermodynamics of linear open quantum walks}

\author{Pedro Linck}
\email{Contact author: pedro.linck@ufpe.br}
\affiliation{Departamento de Física, Centro de Ciências Exatas e da Natureza, Universidade Federal de Pernambuco, Recife-PE, Brazil}

\author{Nadja K. Bernardes}
\affiliation{Departamento de Física, Centro de Ciências Exatas e da Natureza, Universidade Federal de Pernambuco, Recife-PE, Brazil}



\date{\today}

\begin{abstract}
Open quantum systems interact with their environment, leading to nonunitary dynamics. We investigate the thermodynamics of linear Open Quantum Walks (OQWs), a class of quantum walks whose dynamics is entirely driven by the environment. We define an equilibrium temperature, identify a population inversion near a finite critical value of a control parameter, analyze the thermalization process, and develop the statistical mechanics needed to describe the thermodynamical properties of linear OQWs. We also study nonequilibrium thermodynamics by analyzing the time evolution of entropy, energy, and temperature, while providing analytical tools to understand the system's evolution as it converges to the thermalized state. We examine the validity of the second and third laws of thermodynamics in this setting. Finally, we employ these developments to shed light on dissipative quantum computation within the OQW framework.

\end{abstract}

\maketitle

\section{\label{intro} Introduction}
Quantum thermodynamics is a subject that has garnered considerable attention in recent years \cite{romanelli19, romanelli12, liu07, gogolin2016, skrzypczyk14, lindel09}. Quantum systems' equilibration and thermalization processes, as well as their behaviors, are now being well understood \cite{gogolin2016, romanelli12, lindel09, Goold2016}. Temperatures for quantum states that are not thermalized Gibbs states have also been proposed, such as effective temperatures for more general quantum states \cite{lipka23}. Thermodynamics has also found ground in the foundations of quantum computation \cite{Marcel2023, Blok2025, Aamir2025, Bennett1982} and quantum state preparation \cite{korkmaz20}. Another important question that has been studied is how to extract work from quantum systems, its limitations, and how it relates to classical theory \cite{skrzypczyk14, liu07}. Experimental advances in quantum thermodynamics have also been reported \cite{brunelli18, peterson16, Zanin2019, vieira2023, Myers2022}.

 Quantum walks are a type of quantum evolution that takes place on a graph \cite{Ahar93, QWSA}. The vertices of the graph form the basis of possible places for the walker, and the edges give us the information about how the walker can jump from one node to another. Here, we are interested in open quantum walks (OQWs) \cite{Petr12(1), Petr13}, which are a type of non-unitary quantum walk. Open quantum walks with linear topology can be used to perform dissipative quantum computation (DQC) \cite{Petr12, Cirac09}, with established results on their dynamics and computational power \cite{linck25}.

 A general microscopic model has been proposed for OQWs \cite{Petr15}, showing that the dynamics of the walk are closely related to the thermodynamics of the environment, so the development of theoretical tools to describe the statistical mechanics and thermodynamics of open quantum walks offers a framework to elucidate the environmental constraints and possibly understand the design of experimental realizations. 

 The main contribution of this work is to provide analytical tools to study the thermodynamics of a linear model of open quantum walks. We introduce a steady-state temperature in terms of environmental parameters, allowing for the convergence toward the steady state to be interpreted as a thermalization process. We analyze the equilibrium thermodynamics of the model by characterizing the entropy and the Helmholtz free energy, and by identifying a critical environmental parameter that induces population inversion. We further investigate nonequilibrium thermodynamics, determining the onset, duration, and overall profile of the thermalization process through the time evolution of entropy, energy, and temperature. We also quantify the energy required to modify the OQW parameters, which is essential for assessing the limitations of reservoir engineering for dissipative quantum computation within this model. Finally, we prove the validity of the second and third laws of thermodynamics.

 The paper is organized as follows. In Sec.~\ref{preliminaries}, we briefly review the theoretical foundations of OQWs established in the literature. In Sec.~\ref{equilibrium}, we define a temperature for the equilibrium state of linear open quantum walks and develop the corresponding equilibrium statistical mechanics and thermodynamics, relating environmental parameters to the thermodynamic properties of the system. In Sec.~\ref{non-equilibrium}, we study the nonequilibrium thermodynamics of the model: we derive analytical approximations for the onset and completion times of the thermalization process, provide an approximate analytical expression for the entropy during thermalization, and characterize the behavior of temperature throughout the evolution. In Sec.~\ref{heat}, we investigate heat and entropy exchange mechanisms, calculating the heat required to change the walk parameters and analyzing entropy production over time. In Sec.~\ref{sec:dqc}, we show how the formalism developed here can be used to refine previously established results in dissipative quantum computation based on linear open quantum walks. Finally, in Sec.~\ref{sec:conclusion}, we present our conclusions, final remarks, and future perspectives for the results and tools developed in this work.

\section{\label{preliminaries}Open Quantum Walks } 
In this section, we set notation and review established results on open quantum systems and open quantum walks required for our work.

\subsection{Open Quantum Walks}
A quantum walk is a form of quantum evolution on a graph. The vertices define an orthonormal basis for the Hilbert space position (graph), in which the walker resides, and can transition between nodes according to both graph connectivity and the walker's internal state~\cite{QWSA}. The dynamics of a discrete-time quantum walk is given by the repeated application of a single-step evolution rule. Here we focus on open quantum walks (OQWs), a class of quantum walks whose dynamics is entirely induced by the environment.

\par Any quantum evolution can be described by a completely positive and trace-preserving map $\Lambda\colon \mathcal{B}(\mathcal{V}) \to \mathcal{B}(\mathcal{V})$, where $\mathcal{B}(\mathcal{V})$ denotes the algebra of bounded operators in the Hilbert space $\mathcal{V}$~\cite{OQS}. It is well known~\cite{OQS} that if $\dim(\mathcal{V})=d$, then $\Lambda$ admits an operator-sum (Kraus) representation,
\begin{equation}
\Lambda(\rho)=\sum_{i=0}^{m-1} K_i\,\rho\,K_i^\dagger,
\end{equation}
where $m \leq d^2$ and the Kraus operators $\{K_i\}_{i=0}^{m-1}$ satisfy the completeness relation
\begin{equation}
\sum_{i=0}^{m-1} K_i^\dagger K_i = I.
\end{equation}
This representation is not unique.

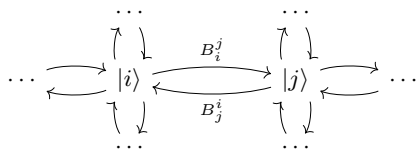
\begin{figure}[b]
\[\begin{tikzcd}[row sep=scriptsize]
	& \cdots && \cdots \\
	\cdots & {\ket{i}} && {\ket{j}} & \cdots \\
	& \cdots && \cdots
	\arrow[curve={height=-6pt}, from=1-2, to=2-2]
	\arrow[curve={height=-6pt}, from=1-4, to=2-4]
	\arrow[curve={height=-6pt}, from=2-1, to=2-2]
	\arrow[curve={height=-6pt}, from=2-2, to=1-2]
	\arrow[curve={height=-6pt}, from=2-2, to=2-1]
	\arrow["{B_i^j}", curve={height=-6pt}, from=2-2, to=2-4]
	\arrow[curve={height=-6pt}, from=2-2, to=3-2]
	\arrow[curve={height=-6pt}, from=2-4, to=1-4]
	\arrow["{B_j^i}", curve={height=-6pt}, from=2-4, to=2-2]
	\arrow[curve={height=-6pt}, from=2-4, to=2-5]
	\arrow[curve={height=-6pt}, from=2-4, to=3-4]
	\arrow[curve={height=-6pt}, from=2-5, to=2-4]
	\arrow[curve={height=-6pt}, from=3-2, to=2-2]
	\arrow[curve={height=-6pt}, from=3-4, to=2-4]
\end{tikzcd}\]\caption{\label{fig:OQW} An arbitrary open quantum walk can be represented by this visual diagram. If there is an omitted edge in a particular diagram, this means that the corresponding operator $B_i^j$ is zero. Figure extracted from Ref.~\cite{linck25}.}
\end{figure}

\par To formally define an open quantum walk, we introduce the walker’s internal Hilbert space $\mathcal{H}$ and the graph (position) Hilbert space $\mathcal{G}$ on which the walk takes place. Let $\{\ket{i}\}_{i\in G}$ denote an orthonormal basis of $\mathcal{G}$, where $G$ is the set of vertices of the underlying graph. For each $i\in G$, we associate a set of linear operators $\{B_i^{\,j}\}_{j\in G}$, with $B_i^{\,j}\colon \mathcal{H}\to\mathcal{H}$, which act on the internal state when a transition from node $i$ to node $j$ occurs. These operators satisfy the completeness condition
\begin{equation}
\label{eqn:kraus_cond_oqw}
    \sum_{j\in G} B_i^{\,j\dagger} B_i^{\,j} = I,
\end{equation}
ensuring trace preservation of the induced dynamics.  To encode the jump together with the corresponding transformation of the internal state, we define the operators
\begin{equation}
M_i^{\,j}=B_i^{\,j}\otimes \ket{j}\!\bra{i}\in \mathcal{B}(\mathcal{H}\otimes \mathcal{G}).
\end{equation}
One can readily verify that Eq.~\ref{eqn:kraus_cond_oqw} implies the completeness relation
\begin{equation}
\sum_{i\in G}\sum_{j\in G} M_i^{\,j\dagger} M_i^{\,j} = I,
\end{equation}
so that the collection $\{M_i^{\,j}\}_{i,j\in G}$ defines a quantum channel. This operator-sum representation specifies an open quantum walk. Pictorially, an OQW may be represented as a directed graph whose edge $i\to j$ is labeled by the operator $B_i^{\,j}$, as can be seen by Fig. \ref{fig:OQW}.  

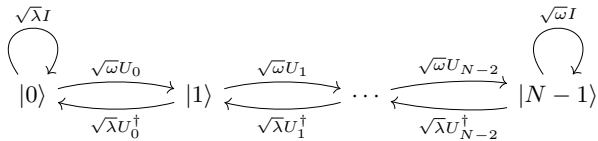
\begin{figure}[h!]
\[\begin{tikzcd}
	{\ket{0}} && {\ket{1}} && \cdots && {\ket{N-1}}
	\arrow["{\sqrt{\lambda} I}", from=1-1, to=1-1, loop, in=55, out=125, distance=10mm]
	\arrow["{\sqrt{\omega} U_0}", curve={height=-6pt}, from=1-1, to=1-3]
	\arrow["{\sqrt{\lambda} U_0^\dagger}", curve={height=-6pt}, from=1-3, to=1-1]
	\arrow["{\sqrt{\omega} U_1}", curve={height=-6pt}, from=1-3, to=1-5]
	\arrow["{\sqrt{\lambda} U_1^\dagger}", curve={height=-6pt}, from=1-5, to=1-3]
	\arrow["{\sqrt{\omega} U_{N-2}}", curve={height=-6pt}, from=1-5, to=1-7]
	\arrow["{\sqrt{\lambda} U_{N-2}^\dagger}", curve={height=-6pt}, from=1-7, to=1-5]
	\arrow["{\sqrt{\omega} I}", from=1-7, to=1-7, loop, in=55, out=125, distance=10mm]
\end{tikzcd}\]
\caption{\label{fig:linearOQW} The diagram corresponding to the linear OQW model. Each $U_i$ is a unitary operator, and $\omega, \lambda \geq 0$ are such that $\omega + \lambda = 1$. Figure extracted from Ref.~\cite{linck25}.}
\end{figure}

Given an initial state $\rho_0\in \mathcal{B}(\mathcal{H}\otimes \mathcal{G})$, the dynamics is defined recursively by
\begin{equation}
\label{eqn:recursive_oqw}
\begin{cases}
\rho^{[0]}=\rho_0,\\
\rho^{[n]}=\displaystyle\sum_{i\in G}\sum_{j\in G} M_i^{\,j}\,\rho^{[n-1]}\,M_i^{\,j\dagger}, \qquad n\ge 1,
\end{cases}
\end{equation}
where $\rho^{[n]}$ denotes the state of the open quantum walk after $n$ steps. It is known~\cite{Petr12(1)} that if the initial state can be written as
\begin{equation}
\rho_0=\sum_{i,j\in G}\rho_{ij}\otimes \ket{j}\!\bra{i},
\end{equation}
for trace-class operators $\rho_{ij}$ satisfying $\sum_{i\in G}\Tr(\rho_{ii})=1$, then after any $n\ge 1$ steps the state $\rho^{[n]}$ becomes block diagonal in the graph basis,
\begin{equation}
\rho^{[n]}=\sum_{i\in G}\rho_{ii}^{[n]}\otimes \ket{i}\!\bra{i},
\end{equation}
where the trace-class operators $\rho_{ii}^{[n]}$ are determined by the recursive evolution in Eq.~\ref{eqn:recursive_oqw}. In particular, after a single step, the coherences between distinct vertices vanish. \\
It is also worth emphasizing that the open quantum walk diagram in Fig.~\ref{fig:OQW}, once supplemented with an initial state, fully specifies the resulting dynamics. For a fixed graph, whenever the directed edge from $\ket{i}$ to $\ket{j}$ is labeled by an operator $C_i^j$, this label can be mapped onto the OQW update rule in Eq.~\ref{eqn:recursive_oqw} through the identification
$M_i^j = C_i^j \otimes \ket{j}\bra{i}$.
Hence, given the graphical description of all edges and their associated operators (with any missing edge understood to correspond to a Kraus operator equal to the zero operator), one may explicitly construct the complete family $\{M_i^j\}$. As an illustrative example, in Fig.~\ref{fig:linearOQW} we obtain
$M_0^0 = \sqrt{\lambda}\, I \otimes \ket{0}\bra{0}$,
$M_0^1 = \sqrt{\omega}\, U_0 \otimes \ket{1}\bra{0}$,
$M_1^0 = \sqrt{\lambda}\, U_0^\dagger \otimes \ket{0}\bra{1}$,
and $M_1^1 = 0$ (for $N>2$ there is no directed edge from $\ket{1}$ to $\ket{1}$), and so on, for a given finite set of unitaries $\{U_i\}$ and non-negative parameters $\omega, \lambda$ satisfying $\omega + \lambda = 1$.

\subsection{Linear Open Quantum Walks}

\begin{figure}[t]
\scalebox{0.6}{\input{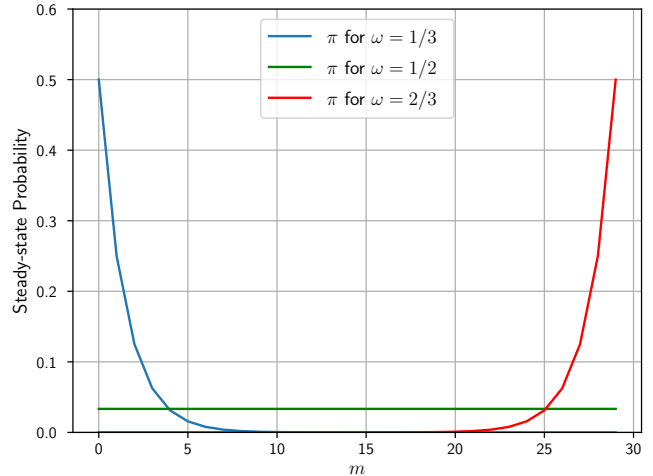}}
\caption{\label{fig:steady_state_omega} Plot of the steady-state probability as a function of the position in the graph for $\omega = 1/3, \; 1/2, \; 2/3$ and fixed $N = 30$.}
\end{figure}

In this work, we consider an open quantum walk on a finite one-dimensional lattice (linear topology), in which each Kraus operator is proportional to a unitary. At each step, the walker moves to the right with probability $\omega$ (or remains at the right boundary) by applying a prescribed unitary, and moves to the left with probability $\lambda$ (or remains at the left boundary) by applying a corresponding unitary chosen to ensure the completeness condition of Eq. \ref{eqn:kraus_cond_oqw}. The graphic representation of the model is shown in Fig.~\ref{fig:linearOQW}. Note that if $\rho_0 = \ket{\psi}\bra{\psi}\otimes \ket{0}\bra{0}$ we have (using the convention $U_{-1} = \mathds{1})$
\begin{equation}
\label{eqn:niteration}
\rho^{[n]} = \sum_i p_{i}^{(n)} U_{i-1}\cdots U_0\ket{\psi}\bra{\psi}U_0^\dagger \cdots U_{i-1}^\dagger \otimes \ket{i}\bra{i},
\end{equation}
where $p_i^{(n)}$ is the probability of finding the walker in node $i$ after $n$ steps \cite{linck25} . This can be seen as a classical Markov chain \cite{linck25} with a transition matrix \cite{MC,SP,HSM}
\begin{equation}
\label{eqn:matrix}
T=\left[\begin{array}{cccccc}
\lambda & \lambda & 0 & & & \\
\omega & 0 & \lambda & &  & \vdots \\
0 & \omega & 0 & & & \vdots \\
\vdots & 0 & \omega & & & \vdots \\
\vdots & & 0 & \ddots & & \vdots \\
 &  &  & & \ddots &  \\
 &  & 
\end{array}\right],
\end{equation}
and the steady-state $\pi$ can be calculated:
\begin{equation}\label{eqn:steady_state}
\pi_m = \dfrac{a^m(a-1)}{a^N-1},
\end{equation}
where $a = \dfrac{\omega}{\lambda} = \dfrac{\omega}{1-\omega}$ and $m$ is a given node in the graph. A plot of the steady-state probabilities for $\omega = 1/3, \; 1/2, \; 2/3$ can be seen in Fig. \ref{fig:steady_state_omega}. Observe that for $\omega < 1/2$ the shape of the steady state is a negative exponential, while for $\omega > 1/2$ it is a positive exponential, and for $\omega = 1/2$ it is the uniform distribution. \begin{review}
For $0<\omega<1$, the chain defined by Eq.~\ref{eqn:matrix} is irreducible, i.e., every node can be reached from any other in a finite number of steps. The chain is also aperiodic, i.e., $\gcd\{n\geq 1 : (T^n)_{m,m}>0\}=1$ for every node $m$, meaning that the walker is not restricted to return to a node only at regular time intervals. Convergence to the unique steady state $\pi$ is therefore guaranteed independently of the initial condition~\cite{linck25, MC}. Moreover, the stationary probability fluxes balance edge by edge,
\begin{equation}
\pi_m T_{m+1,m} = \pi_{m+1} T_{m,m+1}
\quad\Longleftrightarrow\quad
\frac{\pi_{m+1}}{\pi_m} = \frac{\omega}{\lambda} = a ,
\label{eq:detailed_balance}
\end{equation}
which Eq.~\ref{eqn:steady_state} satisfies for every $m$ and every $\omega\in(0,1)$. The steady state is thus reversible, and the relaxation toward $\pi$ can be interpreted as a thermalization process. For $\omega\in\{0,1\}$, the chain is reducible and the walker is absorbed at one of the boundaries.
\end{review}

\section{\label{equilibrium}Equilibrium Thermodynamics of linear Open Quantum Walks}
In this section, we develop the equilibrium thermodynamics of open quantum walks, introducing an equilibrium temperature, deriving the equilibrium statistical mechanics, analyzing the entropy and Helmholtz free energy, and demonstrating a population inversion at $\omega_c = 1/2$. Throughout this paper, we use the von Neumann entropy.

\subsection{\label{subsec:eq_temperature}Thermalization of Linear Open Quantum Walks}
From Eq.~\ref{eqn:steady_state}, we observe that the steady-state probabilities, as a function of the node index $m$, follow an exponential profile $a^{m}=e^{m\log(a)}$. To study equilibrium thermodynamics, we assume that these steady-state probabilities can be interpreted as arising from a Boltzmann distribution $p(m)=e^{-\beta E_m}/Z$. In this way, we have $e^{-\beta E_m} \propto a^m = e^{m \log(a)}$. Therefore, the only way to define an equilibrium temperature for the system in this way is to take $E_m$ to be linear in $m$: $E_m = m \varepsilon + \varepsilon_0$, where $\varepsilon$ is the proportionality factor and $\varepsilon_0$ is a fixed constant. Thus, the system can be interpreted as a harmonic oscillator restricted to a finite number of energy levels, where $\varepsilon$ can be interpreted as the energy gap between consecutive levels and $\varepsilon_0$ as the energy of the ground state, which we set to zero for simplicity. In this way, we obtain $\beta \varepsilon = -\log(a) = \dfrac{\varepsilon}{k_B T}$ and therefore the equilibrium temperature of a linear OQW is given by
\begin{equation}
\label{eqn:temperature}
T = -\dfrac{\varepsilon}{k_B \text{log}(a)} = -\dfrac{\varepsilon}{k_B \text{log}(\omega/\lambda)}.
\end{equation}
Since it is convenient to calculate quantities as functions of $\beta$, it is also useful to note that 
\begin{equation}\label{eqn:omega_beta}
\omega = \frac{1}{1+e^{\beta\varepsilon}} .
\end{equation}
We calculate several statistical-mechanical quantities as functions of $\beta$, but the corresponding statements can always be expressed in terms of $\omega$ using Eq.~\ref{eqn:omega_beta}. Observe that the temperature does not depend on the size of the graph or the initial state of the walker; it only depends on the parameters of the environment $\omega$ and $\lambda$ and the energy gap $\varepsilon$. 

\par It is important to emphasize that we are not deriving a statistical-mechanical description from first principles starting from a microscopic Hamiltonian. Rather, we show that, if the linear OQW model is implemented in a physical system admitting a thermodynamic description, then the energy gaps and temperature must take the form derived here. This follows from requiring the thermodynamic Gibbs equilibrium state to coincide with the mathematical steady state of the corresponding Markov chain. \begin{review}
This identification is supported by the microscopic derivation of OQWs in Ref.~\cite{Petr15}, where the walker interacts weakly with a thermal bosonic bath at inverse temperature $\beta$. For the OQW on a finite chain, whose energy increases linearly with the node index in steps of $\varepsilon$ ($\hbar\Delta_0$ in the notation of Ref.~\cite{Petr15}), the probabilities of hopping to the higher- and lower-energy neighbor are proportional to $\langle n\rangle$ and $\langle n\rangle+1$, respectively, with $\langle n\rangle=(e^{\beta\varepsilon}-1)^{-1}$. Their ratio gives $\omega/\lambda=e^{-\beta\varepsilon}$ in Eq.~(\ref{eq:detailed_balance}), so $\omega$ and $\lambda$ encode the bath temperature and Eq.~(\ref{eqn:steady_state}) is the corresponding Gibbs state.
\end{review}

\par
\begin{figure}[tb]
\scalebox{0.58}{\input{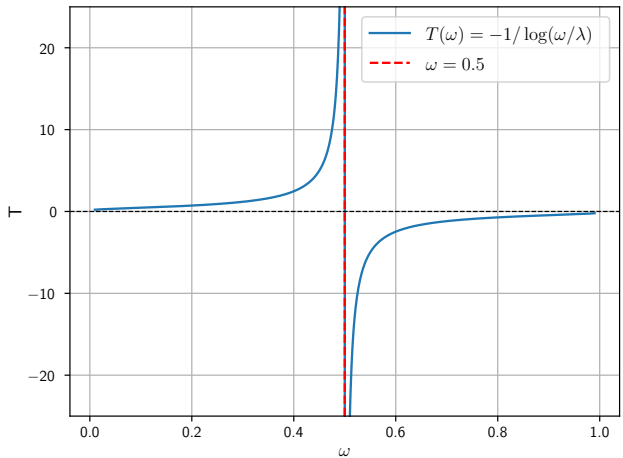}}
\caption{\label{fig:temperature_vs_omega} Temperature of a linear OQW as a function of $\omega$ for $\varepsilon = 1$. The population inversion can be seen as the parameter $\omega$ crosses the divergence in $\omega = 1/2$.}
\end{figure}

 Note that the functional form of the temperature in Eq.~\ref{eqn:temperature} is the same as the initial-state-dependent entanglement temperature for the unitary quantum walks in one dimension \cite{romanelli19, romanelli12}
\begin{equation}
    T_{\text{ent}} = \dfrac{\varepsilon}{k_B \log[\tan(\alpha/2)]},
\end{equation}
where $\alpha$ is a quantity determined by the initial state and $\varepsilon$ is a coupling energy \cite{romanelli12}. The only difference is the parameter inside the $\log$ function, which in the unitary quantum walk involves the initial state, and in our case, requires the environment parameters. The temperature of a linear OQW as a function of $\omega$ is plotted in Fig. \ref{fig:temperature_vs_omega} for $\varepsilon = 1$. There is a discontinuity in temperature at $\omega = \lambda = 1/2$. For $\omega < \lambda$, the temperature remains finite and positive. For $\omega > \lambda$, a population inversion occurs, with higher energy levels becoming more populated than lower ones, leading to a negative temperature. This has the physical meaning that the interaction with the environment maintains the system in an equilibrium steady state. If one tries to extract work from this system, the thermal bath will inject the extracted energy back into the system. It is important to emphasize that we do not claim the physical realization of negative-temperature systems; we only show that our system in this case behaves as if it were thermalizing with a negative-temperature reservoir, providing an effective description. We assume $\varepsilon>0$ throughout the analysis, so that node $\ket{0}$ corresponds to the lower-energy state. The case $\varepsilon<0$ can be handled analogously.

\par A technical point is that we originally defined these quantities only on the graph space. However, since Eq.~\ref{eqn:niteration} shows that the walker and graph spaces are decoupled, we may assign to the state
\[
U_{m-1}\cdots U_0\ket{\psi}\bra{\psi}U_0^\dagger \cdots U_{m-1}^\dagger \otimes \ket{m}\bra{m}
\]
the same energy \(E_m\) that we assign to \(\ket{m}\bra{m}\). Note that the thermodynamic description corresponding to this particular choice of topology and operators for our OQW does not depend explicitly on the walker's internal state space. This is not necessarily the case for more general OQWs.

\subsection{\label{subsec:sm_quantities}Statistical Mechanics Quantities}
From what we have developed in Sec. \ref{subsec:eq_temperature}, we can deduce some statistical mechanics quantities. The partition function for our thermalized linear OQW is given by
\begin{equation}
    Z = \dfrac{a^N-1}{a-1} = \dfrac{e^{-\beta N\varepsilon}-1}{e^{-\beta \varepsilon}-1}.
\end{equation}
The internal energy and its variance are then given by
\begin{equation} \label{eqn:mean_energy}
\langle E \rangle = -\dfrac{\partial \text{log}Z}{\partial \beta} = \frac{\varepsilon}{e^{\beta \varepsilon}-1} - \frac{N\varepsilon}{e^{\beta N \varepsilon}-1},
\end{equation}
\begin{equation} \label{eqn:dispersion_energy}
\langle\Delta E^2 \rangle = \dfrac{\partial^2 \text{log}Z}{\partial \beta^2} = \dfrac{\varepsilon^2 e^{\beta \varepsilon}}{(e^{\beta \varepsilon}-1)^2} - \dfrac{\varepsilon^2 N^2  e^{\beta N \varepsilon}}{(e^{\beta N \varepsilon} - 1)^2}.
\end{equation}
We now calculate the standard deviation $\sigma_E = \sqrt{\langle \Delta E^2\rangle}$. From Eq.~\ref{eqn:dispersion_energy} we obtain, for $N\gg 1$,
\begin{equation}
    \sigma_E = \dfrac{\varepsilon e^{\beta \varepsilon/2}}{|1-e^{\beta \varepsilon}|} = \dfrac{\varepsilon}{2 |\text{sinh}(\beta \varepsilon/2)|}.
\end{equation}
Since we can calculate the entropy as
\begin{equation}
    S(\beta) =  \text{log}Z(\beta) + \beta \langle E \rangle,
\end{equation}
we obtain
\begin{equation}
\label{eqn:entropy_beta}
S(\beta) = \text{log}\left(\dfrac{1-e^{-\beta N \varepsilon}}{1-e^{-\beta \varepsilon}}\right) + \frac{\beta \varepsilon}{e^{\beta \varepsilon}-1} - \frac{\beta N\varepsilon}{e^{\beta N \varepsilon}-1},
\end{equation}
and therefore
\begin{equation}
\label{eqn:derivative_entropy}
\dfrac{\partial S}{\partial \beta} = \beta \dfrac{\partial}{\partial \beta} \langle E \rangle = \dfrac{\beta N^2\varepsilon^2e^{\beta \varepsilon N}}{(e^{\beta N \varepsilon}-1)^2} - \dfrac{\beta \varepsilon^2e^{\beta \varepsilon}}{(e^{\beta \varepsilon}-1)^2}.
\end{equation}
Note that $S(\beta) \to 0$ when $\beta \to \pm \infty$, meaning that the thermodynamics of our system obeys the third law of thermodynamics. The thermodynamic stability of this system follows from:
\begin{equation}
\frac{\partial^2 S}{\partial E^2}
=
\frac{\partial \beta}{\partial E}
=
-\frac{1}{\langle \Delta E^2\rangle}
<0.
\end{equation}
In the thermodynamic limit $N \gg 1$, the first term above vanishes and we obtain
\begin{equation}
\label{eqn:phase_trans_1}
\dfrac{\partial S}{\partial \beta} = - \dfrac{\beta \varepsilon^2e^{\beta \varepsilon}}{(e^{\beta \varepsilon}-1)^2}.
\end{equation}
For $N^{-1} \ll |\beta|\varepsilon \ll 1$, we then have
\begin{equation}
\label{eqn:divergenceentropy}
\frac{\partial S}{\partial\beta} = -\frac{1}{\beta} + \frac{\beta\varepsilon^{2}}{12}.
\end{equation}
A simulation for this large limit is shown in Fig.~\ref{fig:entropy_beta_limit} with $N = 500$. We can see that the entropy is symmetric with respect to the line $\beta = 0$, and the discontinuity of the derivative of the entropy at $\beta = 0$. We also note that at $\beta = 0$, the steady-state distribution is uniform, therefore $S(\beta=0) = \log(N)$, which is the maximum possible value that the entropy can attain in this system. In Fig.~\ref{fig:entropy_beta_limit}, this gives us $S(\beta = 0) = \log(500) \approx 6.21$.

\begin{figure}[h!]
\scalebox{0.58}{\input{figs/S_beta_plot.pgf}}
\caption{\label{fig:entropy_beta_limit} Numerical simulation of entropy versus $\beta$ in the linear OQW for $N = 500$.}
\end{figure}

\par The Helmholtz free energy $F(\beta) = -\dfrac{1}{\beta}\text{log}Z(\beta)$ is
\begin{equation}
F(\beta) = -\dfrac{1}{\beta}\text{log}\left(\dfrac{1-e^{-\beta N\varepsilon}}{1-e^{-\beta \varepsilon}}\right),
\end{equation}
and therefore
\begin{equation}
    \dfrac{\partial F}{\partial \beta } = \dfrac{1}{\beta^2}\text{log}\left(\dfrac{1-e^{-\beta N\varepsilon}}{1-e^{-\beta \varepsilon}}\right) + \dfrac{1}{\beta}\left(\frac{\varepsilon}{e^{\beta \varepsilon}-1} - \frac{N\varepsilon}{e^{\beta N \varepsilon}-1}\right),
\end{equation}
which, in the thermodynamic limit $N \gg 1$, turns into
\begin{equation}
    \dfrac{\partial F}{\partial \beta } = -\dfrac{1}{\beta^2}\text{log}(|1-e^{-\beta \varepsilon}|) + \dfrac{1}{\beta}\dfrac{\varepsilon}{e^{\beta \varepsilon}-1}.
\end{equation}
In the regime $N^{-1} \ll |\beta|\varepsilon \ll 1$ we then have
\begin{equation}
\frac{\partial F}{\partial\beta} = \frac{1-\log|\beta\varepsilon|}{\beta^{2}} + \frac{\varepsilon^{2}}{24} ,
\end{equation}
which also shows a divergence near $\beta = 0$. It is important to note that there is no divergence in the derivatives of $S$ and $F$ for any finite $N$: this divergence only occurs in the large limit $N \gg 1$. We can also analyze the heat capacity 
\begin{equation}
C_V= (\beta \varepsilon)^2 \left[
    \dfrac{e^{\beta \varepsilon}}{\bigl(e^{\beta \varepsilon}-1\bigr)^2}
    - N^2 \dfrac{e^{\beta N \varepsilon}}{\bigl(e^{\beta N \varepsilon}-1\bigr)^2}
\right].
\end{equation}
Observe that for $N \gg 1$,
\begin{equation}
    C_V  = \dfrac{\beta^2 \varepsilon^2 e^{\beta \varepsilon}}{(e^{\beta \varepsilon}-1)^2},
\end{equation}
and for $N^{-1} \ll |\beta|\varepsilon \ll 1$ we obtain
\begin{equation}
C_V = 1 - \frac{(\beta\varepsilon)^2}{12},
\end{equation}
showing that $C_V \to 1$ as $\beta \to 0$ within this window, the analogue of the
Dulong--Petit limit of the Einstein solid ~\cite{Callen1985}.

\subsection{Behavior near $\omega_c = 1/2$}
Let us mathematically analyze the steady-state entropy $S$. The first observation is to prove the symmetry of the steady state under the transformation
\begin{equation}
\begin{aligned}
\omega &\mapsto 1-\omega, \\
m & \mapsto (N-1)-m.
\end{aligned}
\end{equation}

To show this, note that $\omega \mapsto 1-\omega$ is equivalent to $a \mapsto a^{-1}$. Thus,
\begin{equation}
\begin{split}
\frac{a^{-m}(a^{-1}-1)}{a^{-N}-1}
&= \frac{a^{N-m}(a^{-1}-1)}{1-a^N} \\
&= \frac{a^{(N-1)-m}(1-a)}{1-a^N}.
\end{split}
\end{equation}
It follows that
\begin{equation}
S(\omega) = S(1-\omega),
\end{equation}
so that $S(\omega)$ is mirror-symmetric with respect to the line $\omega = 1/2$. This is equivalent to saying that $S(\beta)$ is mirror-symmetric with respect to the line $\beta = 0$. From Eq.~\ref{eqn:divergenceentropy} we observe an abrupt change in the derivative of $S$ near $\beta = 0$, corresponding to a certain critical parameter $\omega_c = 1/2$. The reason for this behavior is that for $\omega_c = 1/2$ and $N\gg1$ the steady-state distribution is uniform and therefore the steady-state probabilities goes to zero everywhere, while for $\omega \neq 1/2$ we have \cite{linck25}
\begin{equation}
    \omega \geq \dfrac{1}{2-\eta} \implies \pi_{N-1} \geq \eta,
\end{equation}
for any chosen $\eta > 0$. Choosing $\eta = 2 - 1/\omega$ for $\omega > 1/2$ leads to $\pi_{N-1} \geq 2 - 1/\omega > 0$ for any $N$. Therefore, regardless of how large $N$ is, there are nodes in the graph for which the steady-state distribution has a positive lower bound independent of $N$. For $\omega < 1/2$, the same occurs for $\pi_0$ by symmetry. This yields two distinct large-$N$ regimes in which the derivative of $S$ is analytic: for $\omega < 1/2$ we obtain a standard Boltzmann distribution with a positive, finite temperature, while for $\omega > 1/2$ we have a population inversion and hence a negative temperature. The same behavior arises when we analyze the free energy $F$, which also exhibits two different classes of analytic behavior in the large-$N$ limit ($N \gg 1$), with a transition at $\omega_c = 1/2$. This behavior highly resembles a phase transition, but can not be characterized as such since it occurs at infinite temperature.

\section{\label{non-equilibrium}Nonequilibrium Thermodynamics of linear open quantum walks}
In this section, we study the nonequilibrium thermodynamics of linear open quantum walks, including the time evolution of the entropy, the thermalization process, and an exponentially decaying law for the entropy. Here, the nonequilibrium regime is defined as the time interval before the system converges to the equilibrium steady state.

\subsection{Thermalization bounds}

\begin{figure}[] 
  \centering
  \begin{subfigure}{\linewidth}
    \centering
    \resizebox{\linewidth}{!}{\input{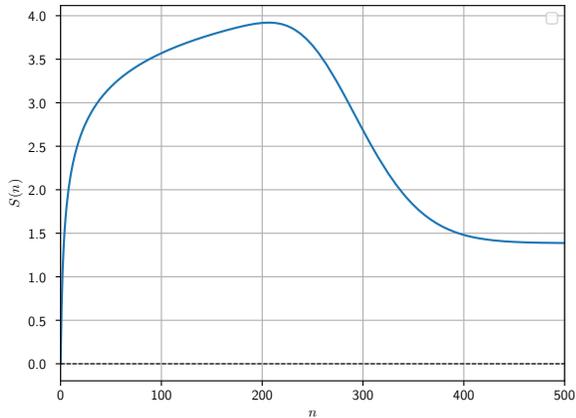}}
    \caption{Numerical simulation of entropy $S$ vs step $n$ for $\omega=2/3,\,N=100$.}
  \end{subfigure}

  \medskip

  \begin{subfigure}{\linewidth}
    \centering
    \resizebox{\linewidth}{!}{\input{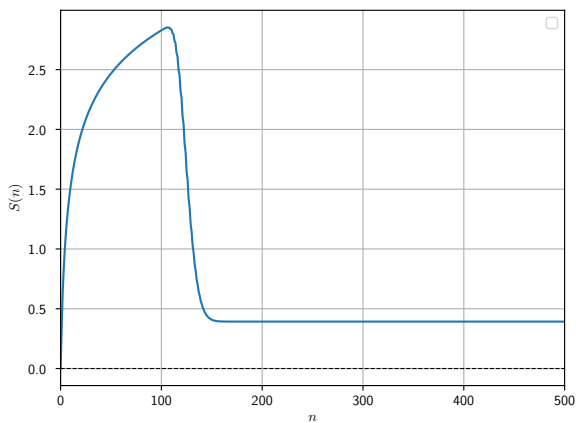}}
    \caption{Numerical simulation of entropy $S$ vs step $n$ for $\omega=9/10,\,N=100$.}
  \end{subfigure}

  \caption{Entropy $S(n)$ of $\rho^{[n]}$ per step $n$.}
  \label{fig:entropy_evolution}
\end{figure}

\begin{figure*}
\scalebox{0.7}{\input{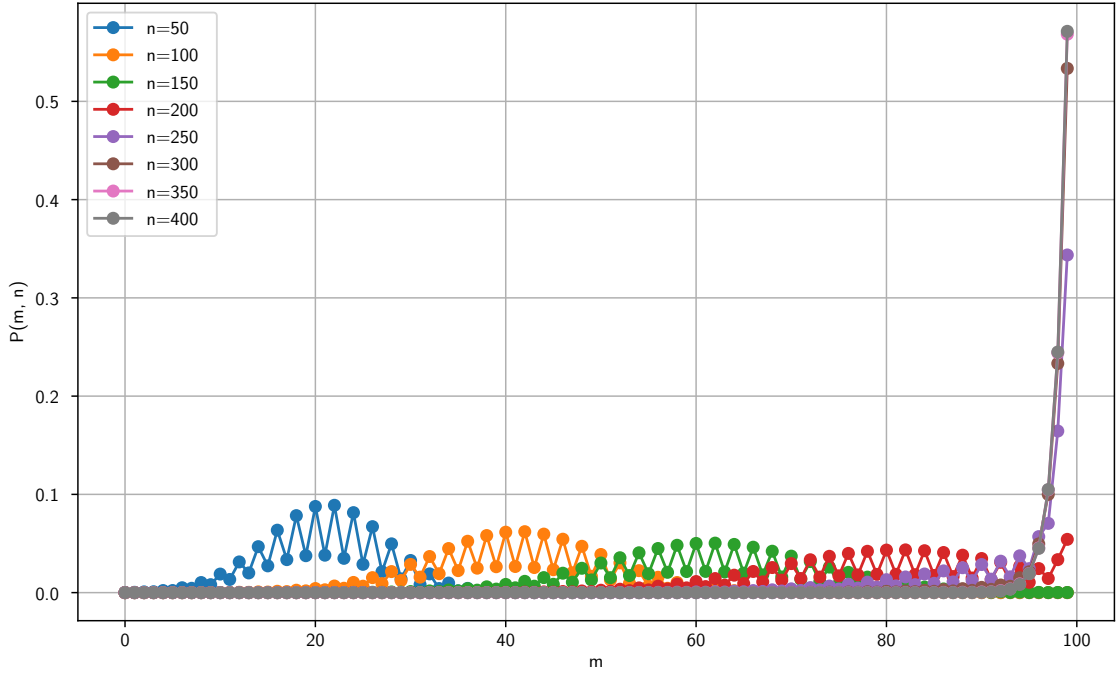}}
\caption{Numerical simulation of the probability evolution of the walker in a linear open quantum walk with $N = 100$, $\omega = 0.7$. For each step $n=50,100,\dots,400$ it is plotted the probability $P(m,n)$ as a function of the node $m$.}
\label{fig:prob_evolution}

\end{figure*}

For now, we assume that the initial state is localized at the first node, which corresponds, in the continuum limit, to a Dirac delta centered at the first node. The generalization of the theory to delocalized pure states follows straightforwardly from the linearity of the evolution. We plot the entropy as a function of $n$ in Fig.~\ref{fig:entropy_evolution} for $N = 100$ and for $\omega = 2/3$ and  $9/10$. The entropy initially grows logarithmically and then decays to the steady-state entropy, indicating that this decaying process is directly related to the thermalization process. What we define as the thermalization process is the process in which the Gaussian distribution starts to deform into the Boltzmann distribution, which is the process of approaching the environment temperature. This deformation process can be seen in Fig.~\ref{fig:prob_evolution}, where we simulate the probability evolution $P(m,n)$ of a walker starting at node $\ket{0}$ with $N = 100$ and $\omega = 2/3$, where $n$ is the number of steps and $m$ is the node.

As shown in Ref.~\cite{linck25}, the probability distribution of this OQW topology, for a suitable initial state, evolves for some time approximately as
\begin{equation}
\label{eqn:probability_approx}
P(x,t) = \frac{1}{\sqrt{4\pi D t}} e^{-(x-vt)^2/4Dt},
\end{equation}
which corresponds to a Gaussian distribution that propagates with velocity $v = 2\omega-1$ and dispersion rate $D = 1/2$. Note that we made the substitution $m \to x$, $n \to t$ and treated them as continuous variables, and note also that this approximation is valid only over a time interval during which the distribution remains sufficiently far from the boundaries. The mean value and standard deviation of $P$ at time $t$ are then $\mu(t) = vt$ and $\sigma(t) = \sqrt{2Dt} = \sqrt{t}$. With these observations, we can argue that the behavior shown in Fig.~\ref{fig:entropy_evolution} is expected. Let us discuss the case $\omega > 1/2$ in detail. First, the distribution behaves approximately as a Gaussian, spreading across the nodes while moving to the right, and therefore its entropy increases. When the distribution reaches the boundary, however, it begins to accumulate there, and the entropy consequently decreases We also note that, in general, the equilibrium state is not a maximum-entropy state along the dynamics. This is a consequence of the open character of the system, which allows for entropy fluxes between the system and its environment.
\\
\\
The time for which the thermalization process starts can be estimated as the time when the position $\mu(t) + 2\sigma(t) = vt + 2\sqrt{t}$ reaches the boundary, i.e., 
\begin{equation}
    vt + 2\sqrt{t} = N,
\end{equation}
which gives us
\begin{equation}
t_{\text{start}} = \left(\frac{\sqrt{1+vN}-1}{v}\right)^2.
\end{equation}
The time in which it stabilizes near the steady-state can be estimated as the time when the position $\mu(t)-2\sigma(t) = vt - 2\sqrt{t}$ reaches the boundary, which similarly gives us
\begin{equation}
t_{\text{end}} = \left(\frac{\sqrt{1+vN}+1}{v}\right)^2.
\end{equation}
Therefore, we find that the time that it takes for the system to thermalize is approximately
\begin{equation}
    t_{\text{therm}} = t_{\text{end}} - t_{\text{start}}= \dfrac{4\sqrt{1+vN}}{v^2}.
\end{equation}
Observe that in the limit $\omega \to 1$, $N>>1$ we have
\begin{equation}
t_{\text{start}} \approx  t_{\text{end}}
\end{equation}
up to terms of order larger than $\mathcal{O}(\sqrt{N})$. For Fig.~\ref{fig:entropy_evolution}(a) we obtain $t_{\text{start}} \approx 213$, $t_{\text{end}}\approx 423$ while in Fig.~\ref{fig:entropy_evolution}(b) we obtain $t_{\text{start}} \approx 100$, $t_{\text{end}} \approx 156$. We draw attention to the fact that this time is closely related to the time required to perform computations in the OQW model for quantum computation \cite{linck25} (see Sec.~\ref{sec:dqc} for further details).

\subsection{Thermalization dynamics}

\begin{figure*}[t]
\scalebox{0.7}{\input{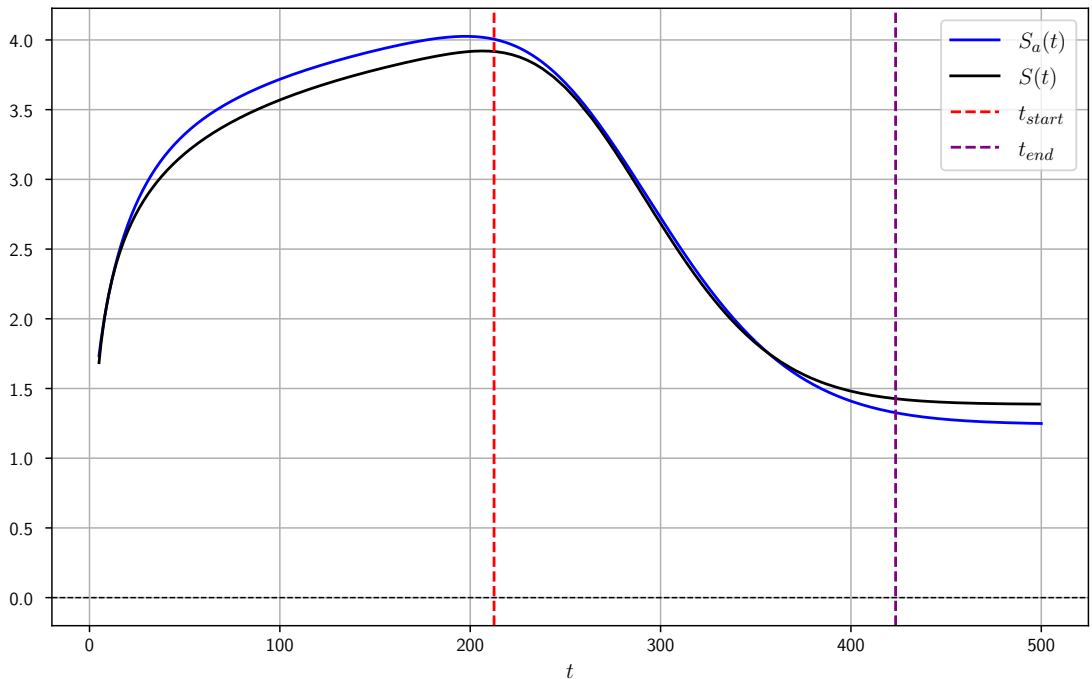}}
\caption{Comparison between the exact entropy $S(t)$ and our analytic approximation $S_a(t)$. Both are plotted as a function of the step $t$. Two vertical lines are drawn corresponding to the thermalization bounds.}
\label{fig:comparison_oqw_entropy}
\end{figure*}

Now we proceed to understand the shape of the entropy as a function of time. We find an approximate analytical formula for the thermalization process, showing an exponential decay until it reaches its equilibrium state, analogous to Newton's law of cooling. The motivation is to understand how fast the system converges after the distribution starts to reach the right boundary From Eq.~\ref{eqn:probability_approx}, it is known, since it is a Gaussian distribution, that
\begin{equation}
    S(t) = \dfrac{1}{2}\log(2\pi e t), \; t < t_{\text{start}},
\end{equation}
which explains the initial logarithmic growth of $S(t)$ in Fig.~\ref{fig:entropy_evolution}. After the thermalization process has started, the normal distribution starts to deform into the Boltzmann distribution of the thermalized state. To calculate an analytical approximation for our entropy in the region where $t \geq t_{\text{start}}$, we propose a heuristic approximation for our probability, supposing it can be divided into two parts, a Gaussian part and a Boltzmann part:
\begin{equation}
    P_a(x,t) = P_G(x,t) + P_B(x,t),
\end{equation}
where $P_a$ is the approximation of the probability distribution and
\begin{equation}
P_G(x,t)=
\begin{cases}
\frac{1}{\sqrt{4\pi D t}} e^{-(x-vt)^2/4Dt}, & \text{if } t \leq N', \\
0, & \text{if } t > N',
\end{cases}
\end{equation}
\begin{equation}
P_B(x,t)=
\begin{cases}
0, & \text{if } t \leq N', \\
\dfrac{1-\text{erf}\left(\frac{N'-tv}{\sqrt{2t}}\right)}{2}\cdot \dfrac{a^x(a-1)}{a^N-1}, & \text{if } t > N',
\end{cases}
\end{equation}
where $N' = N - 2\sigma_{ss}$ with $\sigma_{ss}$ the standard deviation of steady-state probabilities (which can be calculated as $\sigma_E$ putting $\varepsilon = 1$), and the error function is the usual Gaussian error function $\operatorname{erf}(z) = \frac{2}{\sqrt{\pi}}\int_0^{z} e^{-y^2}\,dy$.

We choose to divide the probabilities in the line $x = N'$ because in the steady-state, most of the distribution is concentrated between $N'$ and $N$, which means that the most relevant part of the entropy contribution of $P_B$ is in this region, while this same region is negligible to the entropy contribution of $P_G$. The multiplicative factors in $P_B$ are in order to approximately normalize $P_a$. We do not make an exact normalization to simplify the calculations since we want a simple expression for this approximation. Because $P_a$ is the sum of two probabilities with non-overlapping support, we must have
\begin{equation}
\begin{aligned}
    S_a(t) & = S_G(x,t) + S_B(x,t) \\
    & =  - \bigg( \int P_G(x,t)\log(P_G(x,t))dx \\
    & + \int P_B(x,t)\log(P_B(x,t))dx \bigg),
\end{aligned}
\end{equation}
where $S_a(t)$ is the entropy of $P_a$, which is an approximate expression for the entropy $S(t)$, $S_G$ is the entropy of the Gaussian part of the distribution and $S_B$ is the entropy of the Boltzmann part of the distribution. We can calculate closed expressions for $S_a$ since 
\begin{equation}
\begin{aligned}
S_G(t) & =    - \int_{0}^{N'} P_G(x,t)\log(P_G(x,t))dx \\
 & \approx   - \int_{-\infty}^{N'} P_G(x,t)\log(P_G(x,t))dx \\
  &= \dfrac{(1+\log(2\pi t))\left(1 + \text{erf}\left( \frac{N'-vt}{\sqrt{2t}}\right)\right)}{4} \\
  & - \dfrac{(N'-vt)e^{-\frac{(N'-vt)^2}{2t}}}{2\sqrt{2 \pi t}},
\end{aligned}
\end{equation}
and $S_B$ can be calculated analytically in the same way that we calculated the equilibrium entropy in Sec.~\ref{subsec:sm_quantities}. Note that this expression shows an exponential decay in the gaussian part of the entropy during the thermalization process, which is what we expected from Fig.~\ref{fig:entropy_evolution}.  In Fig.~\ref{fig:comparison_oqw_entropy} we compare our analytical expression $S_a(t)$ obtained above with the exact recursive calculation of $S(t)$ step by step in a linear OQW with $N = 100$ and $\omega = 2/3$. Since the qualitative behavior of the thermalization curve is now well understood, we focus on a more detailed analysis of the data in Tables~\ref{tab:errors_N100} and \ref{tab:errors_N500}, which can be extracted using the codes in the GitHub repository \texttt{OQW-Thermodynamics} \cite{linckOQWThermoRepo}. In Table~\ref{tab:errors_N100} we compare the analytical approximation $S_a(t)$ with the exact recursive entropy $S(t)$ for $N=100$ and $\omega=2/3$. We find that the maximum absolute error between the thermalization bounds is $\delta_{\max}\approx 0.1012$, while the maximum relative error is $\delta_{\mathrm{rel},\max}\approx 0.0709$. The mean relative error is $\overline{\delta}_{\mathrm{rel}}\approx 0.0206$. Moreover, when normalized by the maximum possible entropy $\log N$, the maximum normalized error is $\delta_{\log N,\max}\approx 0.0220$ and the mean normalized error is $\overline{\delta}_{\log N}\approx 0.009$. For higher values of $N$, the analytical approximation becomes increasingly accurate. For example, in Table~\ref{tab:errors_N500} we consider $N=500$ with $\omega=2/3$ and obtain $\delta_{\mathrm{rel},\max}\approx 0.0656$, $\overline{\delta}_{\mathrm{rel}}\approx 0.0160$, $\delta_{\log N,\max}\approx 0.0154$, and $\overline{\delta}_{\log N}\approx 0.0066$, indicating that the approximation is numerically reliable in the limit $N\gg 1$. In principle, improved numerical performance could be achieved by refining the heuristic part of the argument. For example, one may impose that the non-Gaussian part of the distribution is localized between $\langle \pi\rangle-k_1\sigma_{ss}$ and $\langle \pi\rangle+k_2\sigma_{ss}$, with $k_1$ and $k_2$ chosen to optimize the error numerically. Note that $\langle \pi\rangle$ can be computed from $\langle E\rangle$ by setting $\varepsilon=1$.

\begin{table}[t]
\caption{Error metrics for the comparison between $S_a(t)$ and $S(t)$ in a linear OQW with $N=100$ and $\omega=2/3$.}
\label{tab:errors_N100}
\begin{ruledtabular}
\begin{tabular}{lc}
\textbf{Metric} & \textbf{Value} \\
\midrule
$\delta_{\max}$ & $0.1012$ \\
$\delta_{\mathrm{rel},\max}$ & $0.0709$ \\
$\overline{\delta}_{\mathrm{rel}}$ & $0.0206$ \\
$\delta_{\log N,\max}$ & $0.0220$ \\
$\overline{\delta}_{\log N}$ & $0.0090$ \\
\end{tabular}
\end{ruledtabular}
\end{table}

\begin{table}[t]
\caption{Error metrics for the comparison between $S_a(t)$ and $S(t)$ in a linear OQW with $N=500$ and $\omega=2/3$.}
\label{tab:errors_N500}
\begin{ruledtabular}
\begin{tabular}{lc}
\textbf{Metric} & \textbf{Value} \\
\midrule
$\delta_{\max}$ & $0.5386$ \\
$\delta_{\mathrm{rel},\max}$ & $0.0656$ \\
$\overline{\delta}_{\mathrm{rel}}$ & $0.0160$ \\
$\delta_{\log N,\max}$ & $0.0154$ \\
$\overline{\delta}_{\log N}$ & $0.0066$ \\
\end{tabular}
\end{ruledtabular}
\end{table}

\subsection{Temperature in the non-equilibrium regime}
\begin{figure}[t]
\scalebox{0.6}{\input{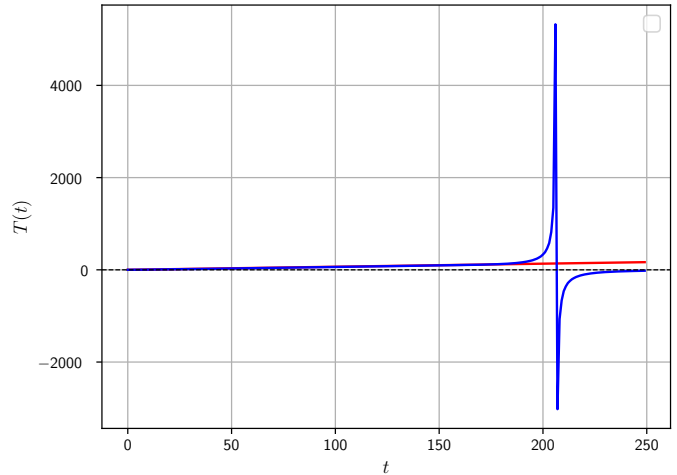}}
\caption{\label{fig:temperature_vs_time} Temperature $T$ as a function of time for $\varepsilon = 1$, $N = 100$, $\omega = 2/3$. The red line is the boundary-free temperature $T_G = 2v\varepsilon t$, the green line is the equilibrium temperature $T_{\text{eq}} = -\varepsilon/\log(\omega/\lambda)$, while the blue curve is the actual temperature of the system.}
\end{figure}

First of all, it is important to note that there is no universal way to define a temperature for nonequilibrium states, and one usually extrapolates some equilibrium relation \cite{CasasVazquez2003}. Here, we follow a trajectory-based definition via the chain rule, similar to one already proposed in Ref.~\cite{Narayanan2012}, and justify our choice based on our calculations and simulations. For a time $t < t_{\text{start}}$, in the regime where the probability distribution is Gaussian, the temperature is a function of time as
\begin{equation}
    T_G(t) = \frac{\partial \langle E \rangle}{\partial S} = \dfrac{d\langle E \rangle/dt}{dS/dt}= 2 v \varepsilon\, t,
\end{equation}
so that, for $\omega > 1/2$ (for example), the temperature starts from zero and increases linearly in time up to a certain point before $t_{\text{start}}$. It is important to note that this linear behavior begins to break down as the distribution deforms toward the equilibrium Boltzmann distribution. A plot for $T(t)$ for $\varepsilon = 1$, $N = 100$, $\omega = 2/3$ can be seen in Fig. \ref{fig:temperature_vs_time}. From the figure, we see that near $t_{\text{start}}$ there is an abrupt temperature change that makes $T$ negative: it is approximately the time at which the distribution starts to accumulate in the last node. In the vicinity of the time at which the thermalization process effectively begins, the temperature rapidly drops to negative values and begins to approach its equilibrium value
\begin{equation}
    T_{\text{eq}} = - \frac{\varepsilon}{\log(\omega/\lambda)}.
\end{equation}
To illustrate the initial behavior, we restrict the graph shown in Fig.~\ref{fig:temperature_vs_time} to the interval $t \in [0,150]$, resulting in Fig.~\ref{fig:temperature_vs_time_initial}. The convergence $T(t) \to T_{\text{eq}}$ can be seen in Fig.~\ref{fig:temperature_vs_time_final}, where $T(t)$ is plotted over the interval $t \in [250,500]$ (blue line) and compared with $T_{\text{eq}}$ (green line). Observe that our choice of effective temperature has physical content away from $t_{\text{start}}$: early in the evolution, while the distribution is Gaussian, the temperature increases, which is compatible with the increasing deviation of the Gaussian distribution and, therefore, with increasing disorder. After $t_{\text{start}}$, the temperature becomes negative, which is compatible with the probabilities accumulating in the higher energy levels, and it exponentially approaches the equilibrium temperature $T_{\text{eq}}$.

\begin{figure}[t]
\scalebox{0.6}{\input{figs/temperature_vs_time_initial.pgf}}
\caption{\label{fig:temperature_vs_time_initial} Temperature $T$ as a function of time for $\varepsilon = 1$, $N = 100$, $\omega = 2/3$ restricted to $t\in [0,150]$ . The red line is the boundary-free temperature $T = 2v\varepsilon t$, while the blue curve is the actual temperature of the system.}
\end{figure}

\begin{figure}[t]
\scalebox{0.6}{\input{figs/temperature_vs_time_final.pgf}}
\caption{\label{fig:temperature_vs_time_final} Temperature $T$ as a function of time for $\varepsilon = 1$, $N = 100$, $\omega = 2/3$ restricted to $t\in [250,500]$ . The green line is the equilibrium temperature $T_{\text{eq}} = -\varepsilon/\log(\omega/\lambda)$, while the blue curve is the actual temperature of the system.}
\end{figure}

\section{\label{heat}Entropy production and heat exchange}
In this section, we determine the energy required to vary the parameters of the walk and analyze the entropy production and the second law of thermodynamics for this system.

\subsection{Energy required to change the parameter $\omega$}
\begin{figure}[b]
\scalebox{0.6}{\input{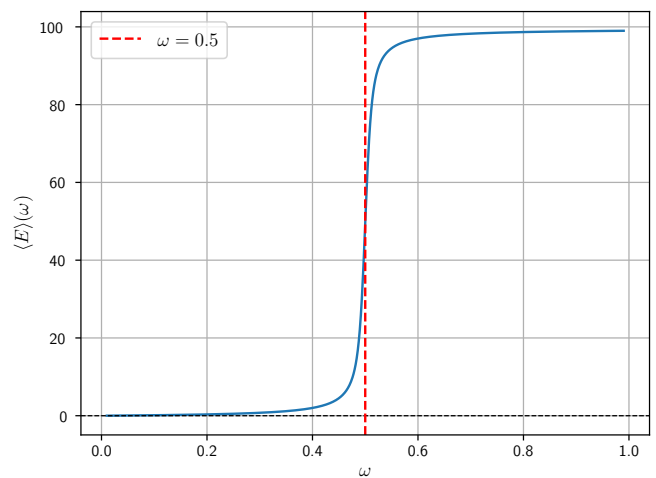}}
\caption{\label{fig:mean_energy_vs_omega} Mean energy $\langle E \rangle$ as a function of the parameter $\omega$ for $\varepsilon = 1$, $N = 100$. Here we have the gap $E_g = 99$.}
\end{figure}

First, we analyze Eq.~\ref{eqn:mean_energy}. For $\beta \to -\infty$, we have $\langle E \rangle \to (N-1)\varepsilon$, while for $\beta \to +\infty$ we have $\langle E \rangle \to 0$. Therefore, the energy gap between the limits $\omega = 0$ and $\omega = 1$ is $E_g = (N-1)\varepsilon$. For large $N$, this gap is concentrated near $\omega_c = 1/2$, where the energy required to change the parameter from $\omega = 1/2 - \delta$ to $\omega = 1/2 + \delta$ becomes approximately $E_g$ for $|\delta| \ll 1$. It is also important to note that changing $\omega$ slightly away from $\omega = 1/2$ requires less energy than changing $\omega$ for values close to $1/2$. A plot of $\langle E \rangle$ as a function of $\omega$ for $N = 100$ is shown in Fig.~\ref{fig:mean_energy_vs_omega}, where we can see the energy gap $E_g = 99$ concentrated near $\omega_c = 1/2$.

If we want to change $\omega \to \omega + d\omega$, the required change in energy is given by the differentiation of Eq. \ref{eqn:mean_energy}:
\begin{equation}
d\langle E \rangle
= \varepsilon \; d\omega\left(
\dfrac{1}{(1-2\omega)^2}
- N^2 \dfrac{(1-\omega)^{N-1}\omega^{N-1}}{\bigl((1-\omega)^N - \omega^N\bigr)^2}
\right),
\end{equation}
which, for large $N$, can be simplified to
\begin{equation}
    d \langle E \rangle = \dfrac{\varepsilon d\omega}{(1-2\omega)^2},
\end{equation}
representing a peak that changes between the different regimes of the parameter $\omega$. We also notice that $d\langle E \rangle$ has the same sign as $d \omega$. Therefore, if we want to increase $\omega$, we need to give energy to the system, and if we want to decrease $\omega$, we have to extract energy from the system. Note that this formalism can therefore be used to estimate the energy required to control the parameter $\omega$ for performing quantum computation (see Sec.~\ref{sec:dqc}).

\subsection{Entropy production}
\begin{figure}[t]
\centering
\scalebox{0.6}{\input{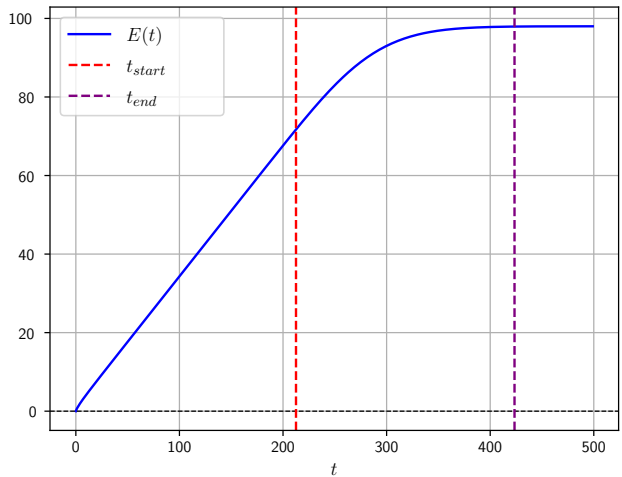}}
\caption{\label{fig:mean_energy_evolution}Time evolution of the mean energy $\langle E\rangle(t)$ for $N = 100$ and $\omega = 2/3$.}
\end{figure}

The internal energy can be estimated analytically using the same techniques developed above to approximate the entropy. In short times, the energy increases approximately linearly in time, since it is proportional to the mean $\mu(t)=vt$ of the underlying Gaussian distribution. After thermalization, it approaches the equilibrium value. A numerical simulation is shown in Fig.~\ref{fig:mean_energy_evolution} for $N=100$ and $\omega=2/3$. The plot displays an initial linear-growth regime, followed by the onset of saturation. The thermalization bounds derived above provide an estimate of when the energy absorption starts to level off. Note that the environment must continuously supply energy to the walk for it to reach the steady state.

The second law of thermodynamics can be written in differential form as $dS = \frac{\delta Q}{T} + \delta S_{\mathrm{gen}}$, $\delta S_{\mathrm{gen}} \geq 0$, where $dS$ is the entropy change of the system, $\delta Q$ is the infinitesimal heat exchanged with the environment at temperature $T$, and $\delta S_{\mathrm{gen}}$ is the entropy generated during the process.

\begin{figure}[b]
\centering
\scalebox{0.6}{\input{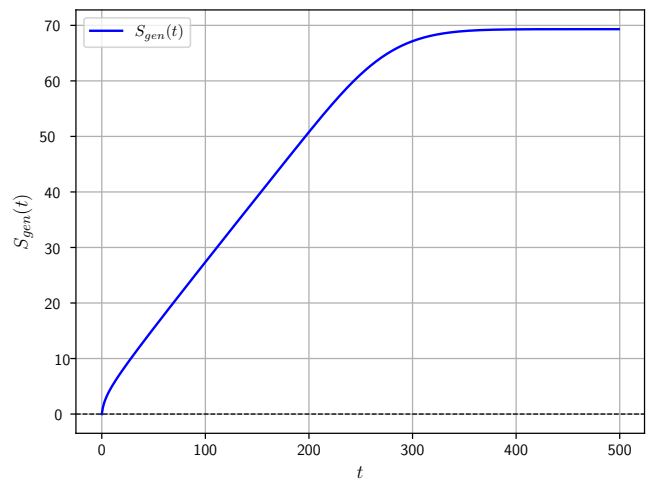}}
\caption{\label{fig:entropy_production}Entropy production $S_{\mathrm{gen}}(t)$ for $N = 100$ and $\omega = 2/3$.}
\end{figure}

In our setting, taking $S(0) = 0$ and $\langle E\rangle(0) = 0$, and assuming a constant bath temperature $T$, we can write the integrated form
\begin{equation}
S(t) = \frac{\Delta Q(t)}{T} + S_{\mathrm{gen}}(t),
\end{equation}
where $S_{\mathrm{gen}}(t)$ is a function that increases monotonically with time and $\Delta Q(t) = \langle E\rangle(t)$ is the total heat absorbed up to time $t$. Since we can compute $S(t)$, $\Delta Q(t)$, and $T$ independently, we can verify the validity of the second law by evaluating $S_{\mathrm{gen}}(t)$ and inspecting its time dependence. In Fig.~\ref{fig:entropy_production}, we calculate $S_{\text{gen}}(t)$ for $N = 100$, $\omega = 2/3$, showing a positive generated entropy for the system, therefore proving the second law of thermodynamics for this system.

We emphasize that the resulting entropy-production curve is qualitatively very similar to that obtained in ~\cite{romanelli19} for closed quantum walks, suggesting a not yet established connection between the thermodynamic behavior of closed and open quantum walk dynamics.

\section{\label{sec:dqc}Application to Dissipative Quantum Computation}
Closed quantum walks are known to be universal for quantum computation \cite{Childs09, Neil10} and can exhibit algorithmic speedups in specific tasks \cite{Qiang24, Venegas12}. In contrast, the open quantum walk framework has been shown to implement dissipative quantum computation efficiently \cite{Petr12}, and the dynamics and computational power of the canonical linear model have been characterized in detail \cite{linck25}. It has also been shown that OQWs can outperform both classical and unitary quantum walks in implementing the PageRank algorithm \cite{Dutta25}. In Ref.~\cite{linck25}, it was shown that the number of steps to perform the computation can be approximated by
\begin{equation}
n_{\text{steps}} = \frac{N}{2\omega-1},
\end{equation}
at which point the probability of finding the walker at the last node becomes computationally feasible. This calculation can be related to the dynamics of thermalization discussed earlier.

Since the computation is detected in the last node, we can estimate two relevant times: the time at which the computation starts to become feasible (which in this paper corresponds to the start of the thermalization process) and the time at which the distribution becomes more concentrated near the boundary (the end of the thermalization process). We showed in Sec.~\ref{non-equilibrium} that the first estimate is given by
\begin{equation}
n_{\text{start}} = \left(\frac{\sqrt{1+vN}-1}{v}\right)^2,
\end{equation}
while the other is
\begin{equation}
n_{\text{end}} = \left(\frac{\sqrt{1+vN}+1}{v}\right)^2.
\end{equation}
Also note that 
\begin{equation}
n_{\text{start}} \leq n_{\text{steps}}\leq n_{\text{end}}.
\end{equation}
For Fig.~\ref{fig:entropy_evolution}(a), for example, we have 
\begin{equation}
    \begin{cases}
        n_{\text{start}} \approx 213, \\
        n_{\text{steps}} \approx 300, \\
        n_{\text{end}} \approx 423.
    \end{cases}
\end{equation}
Therefore, our thermodynamic approach can be used to better control the number of steps required to achieve a higher success probability in fewer steps. We can also use our results from Sec.~\ref{non-equilibrium} to determine the exact energy required to change the success probability of the computation by varying $\omega$, the energy needed to reach the steady state of the computational process, as well as the required entropy.

\section{\label{sec:conclusion} Conclusion}
In this work, we developed analytic tools to describe the equilibrium and non-equilibrium statistical mechanics of linear open quantum walks, as well as their thermodynamic limit. We defined an equilibrium temperature for the system, computed standard thermodynamic quantities such as the entropy and the Helmholtz free energy, and showed that these exhibit a divergence in the thermodynamic limit near the finite parameter value $\omega = 1/2$, corresponding to a population inversion and two different regions of analytical behavior. We concluded that a consistent thermodynamic description is obtained only when the graph space is effectively interpreted as the energy levels of an oscillator. This sheds light on possible experimental implementations of linear OQWs.
\par We also determined the times at which the thermalization process effectively begins and completes and obtained an analytical approximation for the entropy with a small numerical error and the same qualitative behavior as the exact result. This provided a theoretical framework for understanding the dynamics of thermodynamic quantities of the system, such as energy, entropy, and temperature.
\par In addition, we calculated the energy required to modify the parameters of the walk, which is useful for understanding experimental limitations, and showed that the dynamics are consistent with the second law of thermodynamics. In particular, we derived an exact expression for the energy that must be supplied to the environment to start the population inversion.
\par It is also noteworthy that the expression for the equilibrium temperature and the behavior of the generated entropy are similar, in functional form, to their counterparts in the closed quantum walk setting of Ref.~\cite{romanelli19, romanelli12}, even though here the entire walk is immersed in an external environment, whereas in the closed scenario the internal degrees of freedom behave as an open quantum system coupled to the rest of the walk. This suggests a connection between the thermodynamics of closed and open quantum walks.
\par Finally, we plan to extend this analysis to more general OQW topologies and to more general quantum walks with decoherence. In principle, any topology in which all Kraus operators are proportional to unitaries can be treated along similar lines, since the dynamics then admits an underlying classical Markov chain. In such cases, convergence to a steady state can be established under standard conditions such as aperiodicity and irreducibility~\cite{MC}. This, in turn, enables the extraction of an equilibrium thermodynamic description of the system, at least in principle. Even when these convergence conditions fail, one may still seek a non-equilibrium thermodynamic characterization.

\section*{Data availability}
The data and code supporting the findings of this study are publicly available in the GitHub repository \texttt{OQW-Thermodynamics} \cite{linckOQWThermoRepo}.

\begin{acknowledgments}
This study was financed in part by the Coordenação de Aperfeiçoamento de Pessoal de Nível Superior – Brazil (CAPES) – Finance Code 001. N.K.B. acknowledges financial support from CNPq Brazil (442429/2023-1) and FAPESP (Grant 2021/06035-0). The authors acknowledge the financial support of the National Institute of Science and Technology for Applied Quantum Computing through CNPq process No. 408884/2024-0. P.L. is grateful to Allison R. Pessoa for pointing out the thermodynamic similarity
between our model and the Einstein solid.
\end{acknowledgments}

\nocite*
\bibliography{references}

\end{document}